# Simulations of tubulin sheet polymers as possible structural intermediates in microtubule assembly

Zhanghan Wu[1, 2], Hong-Wei Wang[3, 4], Weihua Mu[5, 6], Zhongcan Ouyang[5, 6], Eva Nogales[3, 7], Jianhua Xing[2]*

[1] Program in Genetics, Bioinformatics and Computational Biology, Virginia Polytechnic Institute and State University, Blacksburg, VA 24061

[2] Department of Biological Sciences, Virginia Polytechnic Institute and State University, Blacksburg, VA 24061

[3] Lawrence Berkeley National Laboratory Berkeley, CA 94720

[4] Department of Molecular Biophysics and Biochemistry, Yale University, New Haven, CT 06520

[5] Institute of Theoretical Physics, The Chinese Academy of Sciences, Beijing 100080, China

[6] Center for Advanced Study, Tsinghua University, Beijing 100084, China

[7] Howard Hughes Medical Institute, Department of Molecular and Cell Biology, University of California, Berkeley, CA 94720

---

* Corresponding author. Address: Department of Biological Sciences, Virginia Polytechnic Institute and State University, Blacksburg, VA 24060, U.S.A.,Tel.: (540)231-1359, email: jxing@vt.edu

## *Abstract*

The microtubule assembly process has been extensively studied, but the underlying molecular mechanism remains poorly understood. The structure of an artificially generated sheet polymer that alternates two types of lateral contacts and that directly converts into microtubules, has been proposed to correspond to the intermediate sheet structure observed during microtubule assembly. We have studied the self-assembly process of GMPCPP tubulins into sheet and microtubule structures using thermodynamic analysis and stochastic simulations. With the novel assumptions that tubulins can laterally interact in two different forms, and allosterically affect neighboring lateral interactions, we can explain existing experimental observations. At low temperature, the allosteric effect results in the observed sheet structure with alternating lateral interactions as the thermodynamically most stable form. At normal microtubule assembly temperature, our work indicates that a class of sheet structures resembling those observed at low temperature is transiently trapped as an intermediate during the assembly process. This work may shed light on the tubulin molecular interactions, and the role of sheet formation during microtubule assembly.

## *Introduction*

Microtubules are one of the three major cytoskeleton components in eukaryotic cells [1,2]. They are hollow cylinders consisting of about 13 parallel protofilaments (PF) formed by the head-to-tail assembly of $\alpha\beta$-tubulin heterodimers. Microtubules play important roles in many eukaryotic cellular processes, including intracellular transport, cell motility, mitosis and meiosis. Microtubule dynamic instability, the phenomenon by which a microtubule switches stochastically between assembly and disassembly phases, is known to be a key property for microtubule function. The regulation of microtubule dynamics has been shown to be both of great biological significance during cell division, and of outstanding pharmaceutical value in tumor therapy. For example, Taxol$^©$, the most widely used anticancer agent, targets tubulin and alters microtubule dynamics resulting in mitotic arrest. Therefore, studying the microtubule assembly/disassembly processes is of great relevance for both biological and pharmaceutical purposes.

To explain the process and mechanism of microtubule assembly, various models have been proposed by both experimentalists and theorists [3,4,5,6,7]. In the most simplistic textbook model, during the microtubule assembly process $\alpha\beta$-tubulin heterodimers add one by one onto the growing end of a microtubule. Most of the existing theoretical work is based on this model [4]. However, a number of experimental observations challenge this view. In 1970s Erickson reported an intermediate sheet structure during microtubule assembly (see also Fig. 1a) [6]. He proposed that tubulins first form a two-dimensional open sheet, which in turn closes into tubes (see Fig. 1a). Several other groups observed that fast growth of existing microtubules occurs via the elongation of a gently curved sheet-like structure at the growing end both *in vitro* and *in vivo* [6,7,8]. Using cryo-electron microscopy, Wang and Nogales reconstructed the structure of a curved sheet assembly of GMPCPP-tubulin stabilized by low temperature and high concentration of magnesium [7,8,9]. The use of GMPCPP avoids the complexity due to GTP hydrolysis. This assembly could then directly convert into microtubules by raising the temperature. The authors proposed that it corresponds structurally to the sheet at growing microtubule ends observed by Chrétien and others [7,9]. In this structure tubulin molecules form slightly curved PFs, in the same head-to-tail manner as those in microtubules. However the PFs are paired, with lateral interactions within one pair being indistinguishable from those in microtubules, but with distinct contacts between pairs [5]. Importantly, relative longitudinal displacements between neighboring PFs ("stagger") are the same as in microtubules, indicating that no longitudinal sliding is needed during the sheet-microtubule transition, in agreement with the direct conversion from one to the other. In the remaining of the paper we call this polymer form "ribbon", and reserve the term "sheet" for the observed structure at the end of a growing microtubule. In this work we suggest that the sheet may contain a class of tubulin structures that include the ribbon, all of which contain alternative lateral bonds different from those observed in microtubules. It is important to mention that in the literature the expression "sheet structure" has been used to refer to a protruding end of an incomplete microtubule [4], with no structural difference in the individual dimers or their interactions with respect to that in the microtubule itself, unlike the two-dimensional sheet of Chrétien and coworkers or the stable ribbon assembly of Wang and Nogales.

Wang and Nogales obtained the sheet structure by stabilizing it at low temperatures. An increase in temperature results in the direct conversion of these structures into microtubules. On decreasing the temperature a GMPCPP microtubule converts into the ribbon structure through peeling (Wang and Nogales, unpublished result; also in [10]). This observation implies that the sheet is thermodynamically more stable than the MT at low temperature, but is less stable at higher temperature (Fig. 1b). The conversion resembles a phase transition, which explains the observed sharp temperature dependence [11]. However, the sheet structure is short-lived in conditions under which MTs are formed, suggesting it as a kinetic intermediate [6,7].

The structural observations of Wang and Nogales raised several questions. How can a ribbon structure with alternating lateral interactions be formed during the assembly of tubulins? What is the relation between the ribbon structure and the sheet structures observed at the growing end of a microtubule at physiological conditions? What is the mechanism of the sheet-to-microtubule transition? If the sheet structure is indeed an intermediate in microtubule assembly *in vivo*, is there any biological function for it?

Due to the lack of detailed, atomic formation for the sheet, the ribbon, or the microtubule, as well as detailed kinetic studies, in this work we take an inversed problem approach. First we find out a set of minimal requirements for the system properties to reproduce the experimental observations, specifically the structures of Wang and Nogales. Then we assume that similar properties are applicable to the assembly process at physiological conditions as well, examine the consequent dynamics, and make testable predictions.

## *Methods*

**1. The model**

We assume the $\alpha\beta$-tubulin heterodimer to be the microtubule building block, and neglect direct association/ disassociation of larger filaments, whose contributions are expected to be very small [2]. This assumption is adopted in most existing models. In this work we focus on the GMPCPP tubulins, therefore will not include GTP hydrolysis in the model. We consider three types of reactions (Fig. 2a &b):

1) A dimer can longitudinally add or dissociate from the ends of a PF (Fig. 2a, process 1). The reaction rates for plus and minus ends are different by a constant ratio $\delta$ [12,13]. This ensures that the equilibrium constants are the same for the reactions at both ends, as required by thermodynamics. For convenience in this work we call the noncovalent (longitudinal or lateral) interaction between two tubulins a "bond".

2) A dimer can laterally associate with or dissociate from a PF from either side (Fig. 2a, process 2). The ribbon structure of Wang and Nogales (Fig. 2b) reveals that two neighboring PFs can form two types of lateral bonds [6,7,9]. We call one the *tube* bond as it closely resembles that present in closed, cylindrical microtubules. The other one we called the *sheet* bond, corresponding to that newly observed by Wang and Nogales between PF pairs.

As suggested by our cryo-EM analysis [9], the main sequence regions involved in lateral interactions between PFs in microtubules are the M-loop (Residues 274-286: PVISAEKAYHEQL in $\alpha$-tubulin; PLTSRGSQQYRAL in $\beta$-tubulin) and the N-loop (Residues 52-61: FFSETGAGKH in α-tubulin; YYNEAAGNKY in β-tubulin) [14,15], whereas the lateral sheet bond interactions between two PFs involve site 1 (Residues 336-342 (H10-S9 loop): KTKRTIQ in α-tubulin; QNKNSSY in β-tubulin) and site 2 (Residues 158-164 (H4-S5 loop): SVDYGKK in α-tubulin; REEYPDR in β-tubulin) (Fig. S1a). We identified these stretches of residues based on our low-resolution (18Å) cryo-EM reconstructions, and thus as a coarse approximation to the actual physical interface. Interestingly, the residues involved in the sheet bond are more conserved than those in the tube bond (see Fig. S1b) [16]. It is important to mention that two types of lateral bonds are present in nature in the stable structure of the microtubule doublet, where some PFs need to interact laterally with two neighboring ones simultaneously [17]. The recent doublet structure by Sui and Downing shows a non-MT lateral interaction between PFs B10 and A5 (in their notation) [18]. The doublet and the ribbon structures show that the non-MT interactions in both structures are obtained by rotating one PF relative to another laterally (Fig. S1c). The doublet structure shows even larger rotation angle than the *sheet* bond, possibly further distorted by other binding proteins in this structure [18]. We also noticed that the various structures obtained by Burton and Himes at slightly basic pHs are easily explained by the existence of alternative types of lateral bonds , but molecular details of their structures are lacking [19]. Physically, the existence of two types of lateral bonds means that the potential of mean force between two neighboring tubulin dimers along the lateral rotational angle assumes a double-well shape. This situation is similar to the lateral interactions along the longitudinal direction, where calculations of electrostatic interactions by Sept et al. show a double-well shaped potential, corresponding to the A- and B-typed microtubules [20].

One additional, reasonable assumption is that the formation of the *sheet* bond is dynamically faster than that of the *tube* bond. When two protein molecules (or complexes) encounter each other to form a larger complex, it is unlikely that all the mutual interactions between the two surfaces form all at once. Mostly likely the two protein surfaces form some partial contacts, then gradually adjust to a favorable matching conformation for their mutual interaction, and during the process some residues may need to reorganize slightly. The cryo-EM reconstruction of the low-temperature stabilized ribbons revealed a larger contact surface for the tube bond than for the sheet bond (see Fig. 3). While a larger contact surface may lead to stronger interaction, it may be slower to form. Consequently, a *tube* bond might be slower to form than a *sheet* bond does. However, all these discussions are only suggestive, and further experimental studies are needed. As discussed later, a faster *sheet* bond formation rate is *not* a necessary assumption in our model, but it increases the percentage of transient ribbon structures, and facilitates formation of the sheet structures.

3) We further propose that the two types of lateral bonds can interconvert (Fig. 2b, process 3). Furthermore, two neighboring lateral bonds can mutually affect each other's stability and the inter-conversion rates. This assumption is necessary to reproduce the observed low temperature sheet structure. Physically, it is likely that two consecutive lateral bonds affect each other via allosteric changes in the intervening tubulin molecule. Allosteric effects on the tubulin monomers/dimers have already been proposed to play an important role during the microtubule assembly process, although details are unclear [5,9,21]. For simplicity, in our modeling studies we assume the mutual interaction energy between two *sheet* bonds $\Delta G_{ShSh} > 0$, and other types of interactions $\Delta G_{ShTu} \sim \Delta G_{TuTu} \sim 0$, with *Sh* and *Tu* referring to the *sheet* and *tube* bond, respectively. We will discuss alternative choices later. Below we will show that with these choices one can reproduce the observed low temperature ribbon structure. For a lateral bond conversion reaction, a tubulin dimer needs to rotate about 60 degrees around the longitudinal axis of the neighboring PF [9]. In our simulations of the assembly kinetics and thermodynamic analysis, we do not consider the case in which tubulins within one PF form different types of lateral bonds with their lateral neighbors. Such defects (that tubulins within one PF form different types of lateral bonds with their neighbors) would disrupt the longitudinal and lateral interaction network within the structure, thus be energetically unfavorable, and exist only transiently. This resembles a large class of Ising-type models. For example, protein folding kinetics can often be described by two states without referring to the intermediate transition step. Consequently, our simulation assumes that the tubulin molecules within a PF would rotate collectively and cooperatively. As a consequence, the longer the PF, the harder the rotation is. Also, when a tubulin dimer adds to a PF longitudinally in our kinetic model, it engages in the same lateral bond as the rest of the precedent subunits in the same PF. This approximation greatly simplifies the simulation.

Wang et al. observed the temperature-driven ribbon-microtubule conversion using the GTP analogue GMPCPP [9,11]. Therefore GTP hydrolysis is not a requirement for ribbon/sheet conversion into a microtubule, and thus we did not consider the GTP hydrolysis reaction in this study. We enforce the detailed balance condition by relating the rate constants to the corresponding standard free energy change ($\Delta G^0$). For example, the on rate constant $k_{(+)}$ and off rate constant $k_{(-)}$ for a tubulin addition reaction, is given by [22]

$$\frac{k_{(+)}}{k_{(-)}} = \exp(-\frac{\Delta G^0}{k_B T}), \tag{1}$$

where $k_B$ is Boltzmann's constant, $T$ is the absolute temperature. Following Erickson and others [2,23,24], we divide the standard free energy $\Delta G^0$ into two terms, an entropic energy $\Delta G_{Entropy}$ accounting for the subunit translational and rotational entropic loss due to bond formation—not the overall entropy contribution, and the remaining free energy change $\Delta G_i$. The separation allows proper inclusion of $\Delta G_{Entropy}$ while multiple bonds form simultaneously. For instance, the longitudinal binding/dissociation reaction from the plus (upper) end in Fig. 2a gives

$$\Delta G^0 = \Delta G^0_{long} + \Delta G^0_{Sh} + \Delta G^0_{Tu} - 2\Delta G_{Entropy}, \tag{2}$$

where $\Delta G^0_{long}$ is standard free energy for longitudinal association, $\Delta G^0_{Sh}$ the standard free energy change of forming a *sheet* bond, $\Delta G^0_{Tu}$ the standard free energy change of forming a *tube* bond, and the term $-2\Delta G_{Entropy}$ compensates for overcounting of the entropic free energy loss. Detailed description of the rate constant and entropic term calculations can be found in the supporting text A and B.

## 2. Simulation details

The assembly process was stochastically simulated with the Gillespie algorithm [25]. At each step, we recorded all the species in the system and their numbers. A reaction was randomly selected from a list of all the possible reactions of all the species in the system. We only simulated the early stage of the microtubule assembly process starting from tubulin dimers. All the simulation parameters were provided in Table 1 and figure captions. There are four energy terms in the model. In our simulations, the binding energy for the longitudinal bond $\Delta G_{Long}$, and that of the *tube* lateral bond $\Delta G_{Tu}$, were assigned values -19 $k_BT$ and -15.5 $k_BT$, respectively, close to what used in the literature after taking into account the entropy term $\Delta G_{entropy}$ [2,20,26] (see supporting text B). Currently there is no direct experimental information to determine the values of the other two terms, the sheet-type lateral bond energy $\Delta G_{Sh}$, and the allosteric energy term $\Delta G_{ShSh}$. Instead in this work we will examine how the assembly dynamics is affected by changing the values of these terms. Future experimental results may suggest possible parameter value ranges by comparing with our simulations. All the results reported here were averaged over 60 independent simulations.

In most calculations we used constant free tubulin dimer concentrations. That is, we started the simulations with tubulin dimers only and kept free tubulin dimer concentration at a fixed value throughout the simulations. Experimentally the total tubulin concentration is fixed. However, here we only examine the very early assembly stage where the percentage of tubulins forming assembly clusters is negligible, so the free tubulin concentration is approximately the same as the total tubulin concentration. Using a constant free tubulin concentration provided us the advantage to increase the simulation efficiency with a limited computational resource. It also allowed us to examine the effect of free tubulin concentrations on the assembly process more easily. Exceptions are Fig. 4f, where the total number of tubulin dimers was kept constant, and the results were averaged over 2000 independent simulations. In this case we kept the system in a small size so we could run simulations for a prolonged time until the system reached equilibrium.

At each sampling step, we took a snapshot of the tubulin assembly clusters. Different clusters have different shape, length and width. To characterize the structural properties of each cluster, we examined the following joint probabilities (or percentages): 1) P(Tu-Tu)-- both of the two neighboring lateral bonds lying between three neighboring PFs being *tube* type; 2) P(Tu-Sh)— one tube type, and one sheet type; 3) P(Sh-Sh)-- both being *sheet* type, with P(Tu-Tu) + P(Tu-Sh) + P(Sh-Sh) = 1. We call the local structure formed by three tubulin dimers in lateral contact as a Tu-Tu, Tu-Sh, or Sh-Sh 3-mer structure. The percentage of Tu-Sh structures in the system is calculated as the ratio between the total number of Tu-Sh structures and the total number of 3-dimer structures in all clusters with three or more PFs. A cluster is defined as a ribbon cluster only if P(Tu-Sh) = 1 (Fig. 2c). Therefore a higher value of P(Tu-Sh) means that the cluster is closer to a ribbon structure. A ribbon cluster must have 3 or more PFs by definition. The percentage of ribbon structures in the system is calculated as the ratio between the total number of Tu-Sh structures in the ribbon clusters and the total number of 3-dimer structures in all clusters with three or more PFs. To calculate the population of clusters with certain number ($N$) of PFs, we simply count the total numbers of those $N$-PF clusters at certain steps. The average PF length of an $N$-PF cluster is calculated as the total number of dimers in the cluster divided by $N$.

Currently there is no quantitative experimental data available on the assembly rates at the initial stage we studied here. Therefore all the results are reported with a relative time unit, which can be easily scaled to the experimental rates once available.

## Simulation Results

**1. Effect on tubulin assembly of a difference in binding energy between *sheet-* and *tube-*lateral bonds**

Fig. 4 gives the dependence of the assembly process on the value of the $\Delta G_{Sh} - \Delta G_{Tu}$ (binding energy difference between the *sheet-* and *tube-*type lateral bonds), with fixed values of $\Delta G_{Tu} = -15.5\ k_BT$ and $\Delta G_{Sh} = 6\ k_BT$. The percentage of Sh-Sh structures is negligible for all simulations (data not shown). The percentage of ribbon structures and that of Tu-Sh structure decreases on increasing $\Delta G_{Sh}$ (see Fig. 4a & b). For $\Delta G_{Sh} - \Delta G_{Tu} < 0$ (the *sheet* bond stronger than the *tube* bond, simulating the low-temperature condition) the percentage of ribbon structures stays at a high plateau (top curves in Fig. 4a). For $\Delta G_{Sh} - \Delta G_{Tu} > 0$ (the *tube* bond is stronger than the *sheet* bond, simulating the high-temperature condition) the percentage of ribbon structures starts with a relative high value, then decreases with time. This observation indicates that initially formed *sheet* bonds transform into *tube* bonds at a later time. Fig. 4c supports this idea by showing that (for $\Delta G_{Sh} - \Delta G_{Tu} = 1\ k_BT$) the percentage of Tu-Sh structures in 3-PF clusters is higher than that of later formed larger clusters. Fig. 4d gives (also for $\Delta G_{Sh} - \Delta G_{Tu} = 1\ k_BT$) the average PF lengths (as number of dimers) for different cluster sizes. Small clusters with one or two PFs quickly reach steady-state with average longitudinal length of about 4 tubulin dimers. Experimentally, a large amount of small single- and double-PF clusters with length 4-5 tubulin dimers are observed at the initial stage of the assembly process [11]. The longitudinal length of larger clusters increases continuously within the simulation time. From a thermodynamic point of view the explanation for this result is that the lateral bonds within larger clusters stabilize the clusters, but the single and double-PF clusters lack sufficient lateral bonds and cannot grow long [2]. We performed a simulation with the lateral bond addition turned off so only one PF structures can be formed. The observed average single PF structure length quickly reaches a plateau at a slightly larger value (about 10 dimers, data not shown). From a kinetic point of view, the smaller clusters may disappear also by growing in width and thus transforming into larger clusters before growing long. Similarly shown in Fig. 4e, the populations of single- and double-PF clusters reach a plateau, while the numbers of larger clusters increase continuously within the time of simulation.

In Fig. 4b we examined how the percentage of Tu-Sh structures evolves with time. The results show that all the curves reach a plateau. It is unclear whether the system reaches equilibrium or a dynamic steady-state. The latter would mean that newly formed *sheet* bonds compensate the loss of the Tu -Sh structure population due to Sh→Tu conversion, so the percentage of Tu -Sh structures remains unchanged. If this is the case, the apparent percentage of Sh → Tu conversion should be less than the real value. Therefore, we performed additional simulations with constant total number of tubulin dimers. This time, we used a smaller size system (100 dimers), which allowed us to perform sufficiently long simulations for the system to reach true thermodynamic equilibrium. Fig. 4f shows the evolution of the percentage of Tu-Sh structures with different values of $\Delta G_{Sh} - \Delta G_{Tu}$. In the case of $\Delta G_{Sh} - \Delta G_{Tu} > 0$, thus when the *tube* bond is thermodynamically more stable, the Tu-Sh structures start at relatively high percentage, then convert after the first few thousand steps. This result is due to the faster formation of *sheet* bonds versus *tube* bonds, with the former being transiently trapped as the PFs grow longer. The *sheet* bonds eventually convert to the thermodynamically more stable *tube* bonds and the system reaches equilibrium. Compared to Fig. 4b, we did observe larger percentage of Sh → Tu transition in Fig. 4f, indicating that the curve plateaus in Fig. 4b are due to a dynamic steady-state. In the case of $\Delta G_{Sh} - \Delta G_{Tu} < 0$, where a *sheet* bond is more stable than a *tube* bond, in addition to the effect of the positive $\Delta G_{SS}$, the Tu-Sh structures are more stable thermodynamically (the top lines of Fig. 4f).

## 2. Effect on tubulin assembly of mutual allosteric interaction between two adjacent *sheet* bonds

If formation of a new lateral bond is not affected by the existing PFs ($\Delta G_{ShSh} = 0$), one would expect randomly distributed lateral bond types between PFs. The allosteric term $\Delta G_{ShSh}$ is necessary for reproducing the dominating ribbon structures experimentally observed at low temperature ($\Delta G_{Sh} - \Delta G_{Tu} < 0$). Fig. 5 shows that, for $\Delta G_{Sh} - \Delta G_{Tu} = -1.5\ k_BT$, the percentage of ribbon structures and that of T-S structures is sensitive to the value of $\Delta G_{ShSh}$. As $\Delta G_{ShSh}$ increases from 0 to 6 $k_BT$, the percentage of ribbon structure increases from 20% to around 90% (Fig. 5a). The percentage drops slightly as time evolves. This is because some newly formed small ribbon structures grows to hybrid forms upon adding more PFs. Fig. 5b-d show the 3-mer structure distribution. For $\Delta G_{ShSh} = 0$, Fig. 5b shows that the S-S structure is dominating, reflecting the fact that the *sheet* bond is stronger than the *tube* bond. While there are still about 20% Tu-Sh structures, the Tu-Tu structures are negligible. On increasing $\Delta G_{ShSh}$ to 2 $k_BT$ (Fig. 5c), the free energy difference between a *sheet* and a *tube* bond (-1.5 $k_B$T) cannot compensate the unfavorable term $\Delta G_{ShSh}$, and more Tu-Sh structures than the Sh-Sh structures are formed. As we further increase $\Delta G_{ShSh}$ to 6 $k_BT$ (Fig. 5d), T-S structures become dominating, while the other two structures are negligible. In the case where $\Delta G_{Sh} - \Delta G_{Tu} > 0$, a positive value of $\Delta G_{ShSh}$ maintains its effect on producing higher percentage of newly formed Tu-Sh arrangement, with the ribbon structures dominating the population, but these gradually transform into the more stable microtubule structures (see Fig. S2).

## 3. The effect of free tubulin concentration on the assembly process

The free tubulin concentration is another factor affecting the assembly kinetics. Fig. 6a and b examine the effect of free tubulin concentration on the assembly process in the case where $\Delta G_{Sh} - \Delta G_{Tu} > 0$ (high temperature scenario in which tubulin polymerizes into microtubules). On increasing the free tubulin dimer concentration from 5, to 25, to 125 μM, both the ribbon and T-S structures increase. At higher dimer concentration the population of the ribbon structure forms starts at a high percentage, then drops quickly to the similar level as that at lower dimer concentrations. A possible explanation for this phenomenon is that some of the ribbon structures transform into larger hybrid structures upon PF addition. This is supported by the persistence of the high percentage of Tu-Sh structures at high tubulin concentration (Fig. 6b). The steady-state average length of the single-PF clusters increases as the tubulin concentration goes up (Fig. 6c, curves marked with grey circles), reflecting the fact that increasing the tubulin concentration favors bond formation both thermodynamically and kinetically. The lateral bond formation is apparently favored by high dimer concentrations due to a higher assembly rate, so the multi-PF clusters grow even faster at higher dimer concentration (Fig. 6c, curves marked with open circles). The population of larger clusters (5-PF in the case shown) also increases faster at higher dimer concentrations (Fig. 6d). Overall, our simulations suggest that the sheet intermediates are more likely to be observed at high free tubulin concentrations. This agrees well with the experimental result that larger and more abundant sheet structures are observed during the initial, exponential phase of tubulin of polymerization when free tubulin concentrations are high (>100 µM) [7,27]. Physically, increasing the free dimer concentration increases the cluster growth rates, which effectively allows less time for the internal Sh→Tu transition, and thus increases the percentage of ribbon and Tu-Sh structure, as shown in Fig. 6a & b.

## *Discussion*

Erickson and Pantaloni performed thermodynamic analysis on the initial stages of polymer assembly [24], with the assumption that only one type of lateral bond exists. In their model, the sheet is not structurally different from the microtubule structure. In the present study, and while incorporating recent structural information, we are trying to simulate the very early stages of tubulin polymerization at both low and high (physiological) temperature, making a minimal number of assumptions that will reproduce existing experimental observations. The main conclusions from this exercise follow.

**Thermodynamic analysis:** Let's consider a structure with $2m$ PFs of length *n dimers*. At low temperatures (less than 15 ˚C), the *sheet* bond is more stable than the *tube* bond ($\Delta G_{Sh} - \Delta G_{Tu} < 0$). Therefore, the thermodynamically most stable structure tends to form as many *sheet* bonds as possible. However, the term $\Delta G_{ShSh}$ disfavors a sheet structure with all *sheet* bonds. One can show that, instead, the most stable structure is the one with alternating lateral bonds, provided $\Delta G_{ShSh} > |\Delta G_{Sh} - \Delta G_{Tu}|$. The free energy difference between the structure with neighboring *sheet* bonds and the one with alternating lateral bonds is $n(2m-1)(\Delta G_{ShSh} - \Delta G_{Tu} + \Delta G_{Sh})$. The difference between a *sheet* bond-only structure and an unclosed *tube* bond-only structure is $n(2m-1)(\Delta G_{Sh} - \Delta G_{Tu})$. When $n$ and/or $m$ are large, a small difference in the bond energy leads to a large difference in the Boltzmann weight. The structure with alternating lateral bonds is thus the dominating form. Above a certain temperature, the *tube* bond becomes more stable than the *sheet* bond ($\Delta G_{Sh} - \Delta G_{Tu} > 0$), and the microtubule becomes the most stable polymer form. These thermodynamic considerations explain the results in Fig. 4 and Fig. 5. There are several possible origins on the temperature dependence of $\Delta G_{Sh} - \Delta G_{Tu}$. We discussed them in supporting text C.

For the allosteric effect represented by the term $\Delta G_{ShSh}$, we suggest two possible mechanisms. First lateral interactions have been proposed to straighten a tubulin dimer (this is referred to as the lattice effect) [9,28,29]. Consequently, the lateral interaction surface is in general coupled to straightening, and the allosteric effect proposed here and the lattice effect are closely related and coupled. This effect may exist even if each tubulin monomer is treated as a rigid body. While this is the mechanism we favor, a second alternative scenario is that, as tubulin molecules are flexible, lateral interactions on one side could affect the lateral surface on the other side of the protein.

A sheet structure is a common morphology for biological molecule self-assembly [30,31,32]. Tubulin assembly shares some common features. For example, the ribbon structures are helical, and the tubulins are arranged in a microtubule in a helical manner [9]. Therefore, due to asymmetric off-axis interactions between tubulins these structures are chiral [32]. The general theory discussed by Aggeli et al. may be applied to a more detailed analysis of the tubulin assembly model.

**How is the *sheet* bond kinetically trapped during the assembly process**? At physiological temperatures, where $\Delta G_{Sh} - \Delta G_{Tu} > 0$, the microtubule is thermodynamically at the most stable polymer form. However, Fig. 5 shows that a large population of structures with the *sheet* bonds can still be observed transiently at the initial assembly stage. The steady state population of ribbons will depend on the actual value of $\Delta G_{Sh} - \Delta G_{Tu}$. Fig. 7a schematically illustrates some possible reaction pathways that would lead to a kinetic trap (Fig. 7b). During the early stages of microtubule assembly (which we modeled here), short clusters of a few PFs are assembled. When a dimer adds on to a cluster laterally, it forms a *sheet* bond with a higher probability (1→2) than a *tube* bond (1→3). Thermodynamically the *sheet* bond has the tendency to convert into a *tube* bond, since the *tube* bond has lower free energy (2→3, Fig. 7b, left panel). However, before the slow lateral bond type conversion takes place, another dimer may add on longitudinally at the end of a PF with a higher rate (2→4). Lengthening of the PF further increases the difficulty of lateral bond conversion by increasing the conversion barrier height (4→5, Fig. 7b, right panel). Consequently, the lateral *sheet* bonds are transiently trapped.

The main idea in our proposed mechanism is that there are three major classes of competing processes with different characteristic time scales: longitudinal elongation, lateral association to form a *tube*- or *sheet*- type bond, and Sh→Tu conversion. Only the first two processes depend on the tubulin concentrations. As long as the first two processes (especially longitudinal elongation) are much faster than the conversion rate, kinetically trapped structures containing the *sheet* bonds are observable. In our simulations we used a lateral association rate for the *sheet* bond larger than that for the *tube* bond. From a structural point of view, the GTP-tubulin in solution might have a conformation favoring the formation of lateral *sheet* bond over that of the *tube* bond. The oligomerized tubulin may undergo an induced-fit conformational change during the conversion from the *sheet* bond to the *tube* bond, forming more stable lateral interactions. Keeping all other parameters unchanged (e.g., *ΔG_{ShSh}*) but using the same value of the lateral association rates for the two lateral bond types, our simulations (data not shown) show that the ribbon and other hybrid structures are still observed, but constitute a smaller fraction of the total population. It is important to emphasize that our conclusions are quite insensitive to the model parameters used in this work.

Our model also predicts the existence of some hybrid structures between the sheet and the MT forms, where the lateral bond pattern is not so regular (e.g, some of the structures in Fig. 2d and 2e). The cryo-EM images of Chretien et al. revealed a distribution of the sheet bending angles [7], which may correspond to different sheet structures with different ratios of *sheet* versus *tube* bonds. It is tempting to speculate that at the tip of the growing structure Sh-Tu alternating bonds predominate (see Fig. 4), but as the structure gets closer to the growing microtubule, more and more *sheet* bonds have converted to *tube* bonds, until eventually all lateral contacts are *tube* contacts (an alternative explanation is that at any given point along the length of the sheet, all lateral bonds are the same, but that they change in synchrony along the length, asymptotically reaching that of the *tube* bond when the structure finally closes into a tube). The process of conversion was not covered in the present study, where we focused on the very early stage of the assembly process. In this case the formed structures all have small sizes and therefore the conversion process itself is very fast. Instead, it is the initiation of the conversion that is rate-limiting. To mathematically model the conversion process and the sheet curvatures explicitly at the growing tip of a preformed microtubule, one needs to include more details of the mechano-chemical properties of the system. This is an ongoing effort in our labs.

In our model we choose $\Delta G_{ShTu} \sim \Delta G_{TuTu} \sim 0$, and $\Delta G_{ShSh} > 0$. These are roughly based on steric constraints imposed by the competing strains of two distinct curvatures-- the longitudinal curvature along the length of a protofilament, and the curvature of the lateral interactions that give rise to a close structure for the microtubule. Our model also assumes that the value of $\Delta G_{Sh} - \Delta G_{Tu}$ vary with temperature (Fig. S3a). It is important to point out that this scheme (Scheme 1) is not the only one that can reproduce the observed low and high temperature structures (ribbons and microtubules, respectively, at steady state). For example, an alternative scheme (Scheme 2) could be that $\Delta G_{Sh} - \Delta G_{Tu} > 0$ (so the *tube* bond is always stronger than the *sheet* bond), $\Delta G_{ShTu} \sim \Delta G_{ShSh} \sim 0$ (which are unnecessary but for simplicity), but $\Delta G_{TuTu} > 0$, which decreases with temperature (Fig. S3b). Also see supporting text C, which provides some theoretical analysis with a reaction path Hamiltonian [33] on a possible origin for a hypothetical temperature dependence of $\Delta G_{TuTu}$. Our stochastic simulations confirm that this scheme can reproduce the low temperature ribbon structures and the high temperature transient sheet structures (see Fig. S4 and supporting text C for details). Compared to Scheme 1, which is the focus of this work, and where the Sh-Sh structure is negligible (with $\Delta G_{ShSh} > 0$), Scheme 2 suggests that a larger percentage of Sh-Sh structures should be observable if one chooses $\Delta G_{ShSh} \sim 0$. A specific way to distinguish the two schemes would be to examine the population difference of 2-PF clusters with *sheet* bond and *tube* bond. Fig. S5 shows that, in Scheme 1, the *sheet*-type 2-PF clusters are dominant at low temperature and the *tube*-type 2-PF clusters become more at high temperature; in Scheme 2, the *tube*-type 2-PF clusters are always dominant at both high and low temperature. Experimentally determining the 2-PF cluster structures at both low and high temperatures would allow us to estimate the value of $\Delta G_{ShSh}$. This will also help on evaluating the two schemes discussed here and the proposal by Chrétien as well. However, no matter which scheme is correct, our main conclusion remains: the existence of the sheet tubulin structures is due to thermodynamics at low temperatures, but kinetics at higher (physiological) temperatures.

Fygenson et al. carried out variability-based alignment of α- and β- tubulin sequences [16]. More conserved residues usually have functional importance. In Fig. S1 we reproduced their result, and indicated the above-mentioned residues involved in lateral interactions. It is clear that those residues (especially several charged ones) involved in the *sheet* bond formation are generally more conserved than those for the *tube* bond. In addition, there are a smaller number of residues involved in the interface of the former, which can be visualized in a simple fashion by examination of the ribbon electron density map showing a smaller contact surface for the *sheet* bond than for the *tube* bond (see Fig. 3). These observations may explain why the *sheet* bond would be faster to form than the *tube* bond. The former involves less residues but strong electrostatic interactions, which can guide the approaching tubulins to interact. On the other hand, to form a *tube* bond more residues need to align (and reorganize) properly with each other, which may result in a high barrier for the reaction. Is it possible that tubulin evolved a *sheet*→ *tube*, two-step processes to increase the tubulin lateral assembly rate: a free tubulin dimer would first be captured by the fast-forming *sheet* bond, and this would serve as a primer to guide the complex to form the more stable but slower-to-form *tube* bond. In a direct *tube*-bond formation mechanism, the interaction between the loosely formed contact pairs may be too weak to hold the newly added tubulin dimer for sufficiently long time before necessary conformational reorganization takes place to form the stable *tube* bond, which would result in very low lateral association rate.

**How biologically relevant is the proposed *sheet* bond?** The ribbon structure obtained by Wang and Nogales shows two types of lateral bonds. In our model, we assume that the same types of lateral bonds exist during the assembly process of both GMPCPP and GTP tubulins at physiological conditions. One may argue that the observed ribbon structure is not physiological, as it is obtained at low temperature and high magnesium ion concentrations. High magnesium ions are typically used for the stabilization of all forms of tubulin assembly, and are hypothesized to shield the acidic C-terminal tails of tubulin (E-hooks), perhaps in a manner similar to that proposed for classical MAPS. These MAPs are highly basic, poorly structured, and generally have also a stabilizing effect on different tubulin assembly forms (e.g. they stabilized both microtubules, and tubulin rings). Cold temperature, on the other hand, is known to have a destabilizing effect on microtubules (interestingly, the addition of certain +TIPS –proteins that in the cell bind to the growing end of microtubules– to microtubules in vitro renders the polymers cold-stable, just like the anticancer drug taxol does (K. Patel, R. Heald, and E. Nogales, unpublished results)). The formation of the ribbon structure, in the presence of GMPCPP, at low temperatures, was therefore a surprise. A working hypothesis to explain the assembly of the ribbons, in conditions where GTP tubulin would not be able to assemble into microtubules, is that temperature slows down tubulin interactions, with less of an effect on the rate of hydrolysis once a tubulin-tubulin contact has formed. Thus, under low temperature conditions little assembly occurs, and when it does hydrolysis quickly follows, before tubulin has a chance to make a microtubule closure and store the energy as lattice strain. When the hydrolysis step is eliminated, the slow polymerization of GTP tubulin (GMPCPP) can continue without the conformational change, and the destabilization effect that hydrolysis brings on tubulin. Under this simple assumption, we propose that the ribbon assembly conditions shed information on the process of microtubule assembly taking place before microtubule closure. This idea is supported by the structure of the ribbon itself, which shows alternating lateral contacts between protofilaments, that otherwise preserve the precise stagger between protofilaments seen in the microtubule. This suggests that the ribbons would be able to convert directly into microtubules, as it was experimentally confirmed [9]. Concerning the rotation of the lateral *sheet* bond, it is important to mention that this type of arrangement, or at least one involving alternative lateral contacts without longitudinal displacements between protofilaments, could have been deduced directly from the extended sheets observed by Chretien and colleagues growing at the end of microtubules, unless extreme deformability is otherwise hypothesize for the tubulin subunit, which is beyond reason.

An alternative model for the experimentally observed sheet structures at the end of growing microtubules is that they involved tubulin interactions are not different from those observed in a MT. A sheet structure is simply an incomplete protruding MT structure. However, the stochastic modeling results of VanBuren show that with this model it is very unlikely to form long incomplete structures at a MT growing end. The structures are energetically unfavorable, and are precursors for disassembly rather than assembly [4]. They didn't examine dependence of the sheet length on the tubulin concentrations. One would expect weak or inverse dependence, since low tubulin concentrations would favor disassembly. This is contrary to the observation that the sheet structures under observed under growth conditions, and become longer (up to several hundred nanometers) upon increasing tubulin concentrations [7].

In conclusion, although there is yet no direct evidence of the presence of the sheet-type lateral bond described here under physiological conditions (the transient character preventing structural characterization, but see discussions on the doublet below), there is very compelling evidence that alternative lateral interactions do exist in a transient intermediate, the sheets at the end of fast growing microtubules. All our analyses indicate that the ribbon structure is the best candidate in existence to describe such intermediates. A somehow similar, and stable structure has been observed in the doublet form, which demonstrates that the alternative lateral bonds do exist *in vivo*. As discussed below, the unusual high conservation of the residues proposed to participate the *sheet* bond formation strongly suggest the functional importance of these residues. We put forward the proposal that existence of (at least) two types of lateral bond naturally explains the sheet and microtubule forms observed *in vitro*, and the interconversion between them.

The situation *in vivo* is more complex, where various microtubule-associated-proteins (MAPs) may modify the microtubule assembly/disassembly process. While more *in vivo* studies are necessary to address the functional relevance of the sheet structure observed *in vitro*, it will also be very informative to study the microtubule assembly process in the presence of purified microtubule-binding proteins. It is important to notice that all structural studies of microtubules with binding partners have been carried out by adding the partners to preassembled (usually taxol-stabilized) microtubules. The effect on the assembly process of +TIPs, for example, should come a lot closer to reproduce what goes on inside the cells, than the analyses carried out to date with purified tubulin alone.

We also suggest that the existence of alternative lateral bond types may have functional importance. Nogales and Wang proposed that the ribbon structure (and the sheet structure in general) could provide a novel surface for microtubule-binding proteins that could recognize surfaces unique to the *sheet* bond to track microtubule growing ends [5]. It has also been proposed that the sheet structure could constitute a structural cap at the end of growing microtubules [7] of essential importance in dynamic instability (notice that both functions would most likely be linked). Additionally, if the MT lateral bond is indeed stronger than the *sheet* lateral bonds, free energy would be stored in the lateral bonds of the sheet structure, released upon closure, which could result in mechanical force generation. We provided a more detailed discussion of this idea in the supporting text.

The nature of lateral interactions also affects the microtubule mechanical properties. Even if only one type of tubulin lateral interactions exists under normal conditions, microtubules in a cell are constantly under mechanical stress due to protein motors and other microtubule associated proteins [34]. There is a certain probability that some of the lateral bonds within a microtubule may convert to another type of interactions under extreme conditions (e.g. buckling under large mechanical force), as implied by recent atomic force microscope studies [35,36]. The new type of lateral bond provides a way of releasing local mechanical stress without breaking the MT. We expect that the mechanical property of a MT with and without this new type of lateral interaction would be dramatically different, and can be tested experimentally. It remains to be examined if these conditions are biologically relevant.

**What could be the function of the sheet intermediate?** In addition to the artificially generated ribbon structure of Wang and Nogales, cryo-EM studies have more directly shown the presence of sheet intermediates during microtubule growth, both *in vitro [6,7] and in vivo* [8]. Theoretically, the *sheet* structures and the conversion into microtubules could play several important functional roles. Nogales and Wang proposed that the ribbon structure (and the sheet structure in general) could provide a novel surface for microtubule-binding proteins that could recognize surfaces unique to the *sheet* bond to track microtubule growing ends [5]. It has also been proposed that the sheet structure could constitute a structural cap at the end of growing microtubules [7] of essential importance in dynamic instability (notice that both functions would most likely be linked) .

The *sheet* bond involves fewer residues but strong electrostatic interactions, which can guide the approaching tubulins to interact. On the other hand, to form a *tube* bond more residues need to align (and reorganize) properly with each other, which may result in a high barrier for the reaction. Is it possible that tubulin evolved a sheet → tube, two-step processes to increase the tubulin lateral assembly rate: a free tubulin dimer would first be captured by the fast-forming sheet bond, and this would give the formed cluster longer time to adjust to the more stable but slower-to-form tube bond. In a direct tube-bond formation mechanism, the interaction between the loosely formed contact pairs may be too weak to hold the newly added tubulin dimer for sufficiently long time before necessary conformational reorganization takes place to form the stable tube bond, which would result in very low lateral association rate.

We would like to propose here that there could be also a mechanical function for a preformed sheet that eventually closes into microtubule structure. Terrell Hill first proposed that assembly and disassembly of cytoskeletal filaments could generate mechanical force [37]. Subsequent theoretical studies and experimental measurements confirmed this idea [38,39,40,41]. Oster and coworkers proposed a ratchet mechanism and its variations to explain how elongating polymers like microtubules can generate force and push an object forward (see Fig. 8a) [42,43]. Thermal motions of the object and the polymer can produce space between them sufficiently large for a building unit (a tubulin dimer in this case) to add to the polymer's end. Addition of the new unit prevents the object from moving back. Therefore, the random thermal motion of the object is ratcheted into directional motion at the expense of free energy released from unit addition. Most published work uses the ratchet model to explain force measurements during microtubule assembly [39,44]. With the sheet intermediate, the ratchet effect can generate force at the growing tip or at the zipping front, depending on the location of the load. Interestingly, it could also provide another active force generating mechanism in addition to the passive ratchet model. If the MT lateral bond is indeed stronger than the *sheet* lateral bonds, free energy would be stored in the lateral bonds of the sheet structure. Transformation to the MT structure is a cooperative process. When many lateral bonds transform together, they would release free energy much larger than that stored in a single lateral bond, thus enable them to push against larger loads. (Fig. 8b) In this way the energy accumulation step (tubulin bond formation) and the work-performing step (*tube* closure) are temporally and spatially separated. A similar mechanism of performing mechanical work using prestored energy has been proposed for the extension of the Limulus polyphemus sperm actin bundle [45] . Which mechanism dominates would depend on where the contact point between the MT and the load is and on the free energy difference between two types of lateral bonds.

## *Conclusion and future work*

In this study, using the single assumption that there are nearest-neighbor interactions between two consecutive PFs, together with existing structural information, we were able to generate a simple model to explain a large number of observations concerning the mechanism of microtubule assembly. We suggest that the sheet structure observed during microtubule growth may be a kinetically trapped intermediate, and that it is related to the ribbon structure stabilized at low temperature. Our model predicts that the sheet structures are more likely to be observed at high free tubulin concentrations. Structural studies of 2-PF clusters during the assembly process could provide information to discriminate among several possible mechanistic schemes.

Our current analysis has focused only on the initial stage of *in vitro* microtubule assembly. A future study should provide a more detailed description of the assembly process, especially the interface between the sheet bonds and the tube bonds along the longitudinal direction within the growing end of a microtubule. Our current treatment that all the lateral bonds within a pair of PFs are identical is clearly only an approximation. In this work we focused on the assembly dynamics of GMPCPP tubulins. We didn't include GTP hydrolysis dynamics and the resulting tubulin dimer conformational changes. We assume that the structural information extracted from the GMPCPP sheet structure can be extrapolated to the normal assembly process. While supported by several other independent experimental evidences, this assumption requires further scrutiny. Especially we propose that at physiological conditions tubulins can form alternative lateral bond type other than that observed in microtubules, as evidenced in the doublet structure. If being confirmed, it would greatly modify our understanding on the mechanical properties of microtubules, and possible mechanisms of interactions between microtubules and microtubule association proteins (MAP) [2,34,46].

Our current model is essentially a two-dimensional model. The current simple model already provides many new insights on the very initial stage of the assembly process with only small cluster structures formed. Both the sheet and the MT forms are actually three-dimensional manifolds. More structural details are needed to fully account for the helical shape of the sheet and the microtubule structure. In the future a three-dimensional mechano-chemistry model parallel to what have been developed for the direct dimer-addition model would be needed [3,4].

## *Acknowledgements*

We thank Drs Haixin Sui and Ken Downing for providing the doublet structure, and Dr Jian Liu for reading the manuscript and providing helpful comments. This work was supported by a grant from the National Institute of General Medical Sciences of the US National Institutes of Health (E.N), and by a Biomedicine chair from the BBVA Foundation (E.N). E.N. is a Howard Hughes Medical Institute Investigator.

## *References*

1. Alberts B, Johnson A, Lewis J, Raff M, Roberts K, et al. (2002) Molecular Biology of the Cell. New York: Garland.
2. Howard J (2001) Mechanics of Motor Proteins and the Cytoskeleton. Sunderland, MA: Sinauer.
3. Molodtsov MI, Ermakova EA, Shnol EE, Grishchuk EL, McIntosh JR, et al. (2005) A molecular-mechanical model of the microtubule. Biophys J 88: 3167-3179.
4. VanBuren V, Cassimeris L, Odde DJ (2005) Mechanochemical Model of Microtubule Structure and Self-Assembly Kinetics. Biophys J 89: 2911-2926.
5. Nogales E, Wang HW (2006) Structural intermediates in microtubule assembly and disassembly: how and why? Curr Op Cell Biol 18: 179-184.
6. Erickson HP (1974) Microtubule surface lattice amd subunit structure and observations on reassembly. J Cell Biol 60: 153-167.
7. Chretien D, Fuller SD, Karsenti E (1995) Structure of Growing Microtubule Ends - 2-Dimensional Sheets Close into Tubes at Variable Rates. J Cell Biol 129: 1311-1328.
8. McIntosh JR, Grishchuk EL, Morphew MK, Efremov AK, Zhudenkov K, et al. (2008) Fibrils Connect Microtubule Tips with Kinetochores: A Mechanism to Couple Tubulin Dynamics to Chromosome Motion. Cell 135: 322-333.
9. Wang H-W, Nogales E (2005) Nucleotide-dependent bending flexibility of tubulin regulates microtubule assembly. Nature 435: 911-915.
10. Müller-Reichert T, Chrétien D, Severin F, Hyman AA (1998) Structural changes at microtubule ends accompanying GTP hydrolysis: Information from a slowly hydrolyzable analogue of GTP, guanylyl ($\alpha$, $\beta$)methylenediphosphonate. Proc Natl Acad Sci U S A 95: 3661-3666.
11. Wang HW, Long S, Finley KR, Nogales E (2005) Assembly of GMPCPP-bound tubulin into helical ribbons and tubes and effect of colchicine. Cell Cycle 4: 1157-1160.
12. Summers K, Kirschner MW (1979) Characteristics of the polar assembly and disassembly of microtubules observed in vitro by darkfield light microscopy. J Cell Biol 83: 205-217.
13. Bergen LG, Borisy GG (1980) Head-to-tail polymerization of microtubules in vitro. Electron microscope analysis of seeded assembly. J Cell Biol 84: 141-150.
14. Nogales E, Whittaker M, Milligan RA, Downing KH (1999) High-Resolution Model of the Microtubule. Cell 96: 79-88.
15. Li H, DeRosier DJ, Nicholson WV, Nogales E, Downing KH (2002) Microtubule Structure at 8 Å Resolution. Structure 10: 1317-1328.
16. Fygensonm D, Needleman D, Sneppen K (2004) Variability-based sequence alignment identifies residues responsible for functional differences in $\alpha$ and $\beta$ tubulin. Protein Sci 13: 25-31.
17. Amos L, Klug A (1974) Arrangement of subunits in flagellar microtubules. J Cell Sci 14: 523-549.
18. Sui H, Downing KH (2006) Molecular architecture of axonemal microtubule doublets revealed by cryo-electron tomography. Nature 442: 475-478.

19. Burton PR, Himes RH (1978) Electron microscope studies of pH effects on assembly of tubulin free of associated proteins. Delineation of substructure by tannic acid staining. J Cell Biol 77: 120-133.
20. Sept D, Baker NA, McCammon JA (2003) The physical basis of microtubule structure and stability. Pro Sci 12: 2257.
21. Rice LM, Montabana EA, Agard DA (2008) The lattice as allosteric effector: structural studies of alphabeta- and gamma-tubulin clarify the role of GTP in microtubule assembly. Proc Natl Acad Sci U S A 105: 5378-5383.
22. Hill TL (1985) Theoretical problems related to the attachment of microtubules to kinetochores. Proc Natl Acad Sci U S A 82: 4404-4408.
23. Erickson HP (1989) Co-operativity in protein-protein association : The structure and stability of the actin filament. J Mol Biol 206: 465-474.
24. Erickson HP, Pantaloni D (1981) The role of subunit entropy in cooperative assembly. Nucleation of microtubules and other two-dimensional polymers. Biophys J 34: 293-309.
25. Gillespie DT (1977) Exact Stochastic Simulation of Coupled Chemical Reactions. The Journal of Physical Chemistry 61: 2340.
26. VanBuren V, Odde DJ, Cassimeris L (2002) Estimates of lateral and longitudinal bond energies within the microtubule lattice. Proc Natl Acad Sci U S A 99: 6035-6040.
27. Vitre B, Coquelle FM, Heichette C, Garnier C, Chretien D, et al. (2008) EB1 regulates microtubule dynamics and tubulin sheet closure in vitro. Nat Cell Biol 10: 415-421.
28. Rice LM, Montabana EA, Agard DA (2008) The lattice as allosteric effector: Structural studies of {alpha}{beta}- and {gamma}-tubulin clarify the role of GTP in microtubule assembly. Proc Natl Acad Sci USA 105: 5378-5383.
29. Buey RM, Calvo E, Barasoain I, Pineda O, Edler MC, et al. (2007) Cyclostreptin binds covalently to microtubule pores and lumenal taxoid binding sites. Nat Chem Biol 3: 117-125.
30. Xu H, Wang J, Han S, Wang J, Yu D, et al. (2009) Hydrophobic-Region-Induced Transitions in Self-Assembled Peptide Nanostructures. Langmuir 25: 4115-4123.
31. O'Brien ET, Falvo MR, Millard D, Eastwood B, Taylor RM, et al. (2008) Ultrathin self-assembled fibrin sheets. Proc Natl Acad Sci U S A 105: 19438-19443.
32. Aggeli A, Nyrkova IA, Bell M, Harding R, Carrick L, et al. (2001) Hierarchical self-assembly of chiral rod-like molecules as a model for peptide beta-sheet tapes, ribbons, fibrils, and fibers. Proc Natl Acad Sci U S A 98: 11857-11862.
33. Miller WH, Handy NC, Adams JE (1980) Reaction path Hamiltonian for polyatomic molecules. J Chem Phys 72: 99-112.
34. Brangwynne CP, MacKintosh FC, Weitz DA (2007) Force fluctuations and polymerization dynamics of intracellular microtubules. Proc Natl Acad Sci USA 104: 16128-16133.
35. Schaap IAT, Carrasco C, de Pablo PJ, MacKintosh FC, Schmidt CF (2006) Elastic Response, Buckling, and Instability of Microtubules under Radial Indentation.  91: 1521-1531.
36. de Pablo PJ, Schaap IAT, MacKintosh FC, Schmidt CF (2003) Deformation and Collapse of Microtubules on the Nanometer Scale. Phys Rev Lett 91: 098101.
37. Hill TL (1981) Microfilament or microtubule assembly or disassembly against a force. Proc Natl Acad Sci U S A 78: 5613-5617.
38. Dogterom M, Yurke B (1997) Measurement of the Force-Velocity Relation for Growing Microtubules. Science 278: 856-860.

39. Schek HT, 3rd, Gardner MK, Cheng J, Odde DJ, Hunt AJ (2007) Microtubule assembly dynamics at the nanoscale. Curr Biol 17: 1445-1455.
40. Molodtsov MI, Grishchuk EL, Efremov AK, McIntosh JR, Ataullakhanov FI (2005) Force production by depolymerizing microtubules: a theoretical study. Proc Natl Acad Sci US A 102: 4353-4358.
41. Kerssemakers JWJ, Laura Munteanu E, Laan L, Noetzel TL, Janson ME, et al. (2006) Assembly dynamics of microtubules at molecular resolution. Nature 442: 709-712.
42. Peskin CS, Odell GM, Oster GF (1993) Cellular motions and thermal fluctuations: the Brownian ratchet. Biophys J 65: 316-324.
43. Mogilner A, Oster G (1999) The polymerization ratchet model explains the force-velocity relation for growing microtubules. Eur Biophys J 28: 235-242.
44. Gardner MK, Odde DJ (2006) Modeling of chromosome motility during mitosis. Curr Opin Cell Biol 18: 639-647.
45. Shin JH, Mahadevan L, Waller GS, Langsetmo K, Matsudaira P (2003) Stored elastic energy powers the 60-{micro}m extension of the Limulus polyphemus sperm actin bundle. J Cell Biol 162: 1183-1188.
46. Odde DJ, Ma L, Briggs AH, DeMarco A, Kirschner MW (1999) Microtubule bending and breaking in living fibroblast cells. J Cell Sci 112 ( Pt 19): 3283-3288.
47. Koren R, Hammes GG (1976) A kinetic study of protein-protein interactions. Biochemistry 15: 1165-1171.
48. Northrup SH, Erickson HP (1992) Kinetics of protein-protein association explained by Brownian dynamics computer simulation. Proc Natl Acad Sci U S A 89: 3338-3342.
49. Martin SR, Schilstra MJ, Bayley PM (1993) Dynamic instability of microtubules: Monte Carlo simulation and application to different types of microtubule lattice. Biophys J 65: 578-596.
50. Bayley P, Schilstra M, Martin S (1989) A lateral cap model of microtubule dynamic instability. FEBS Lett 259: 181-184.

*Figure Legends*

**Figure 1. Structural model of the microtubule self-assembly pathway.** (a) Simplified representation of a sheet intermediate and its conversion into a microtubule based on cryo-EM observation of sheets at the end of fast growing microtubules [7] and the structure of the low-temperature stabilized ribbons by Wang and Nogales [9]. (b) Schematic illustration of the idea that the ribbon structure is thermodynamically more stable than the microtubule structure at low temperature (left), but less stable at the physiological temperature where microtubule assembly takes place(right). We proposed that tubulin sheet structures are kinetically trapped intermediates.

**Figure 2. Schematic illustration of the basic concepts in the proposed model of tubulin self-assembly.** (a) Three types of reactions are being modeled: longitudinal (1) and lateral (2) association/disassociation, and (b) the switch between the *tube* and sheet types of bond (3). Blue lines correspond to the *tube* bond and red lines to the sheet bond. The EM-based structures at the top of (b) show the difference between two lateral bond types [9]. (c) A typical ribbon structure with alternating lateral bonds. (d) A typical hybrid structure with the two types of lateral bonds randomly distributed. (e) An end-on view of several possible 5-PF structures.

**Figure 3. Course inspection of the electron density map of the ribbon structure.** It reveals a clearly larger buried interface for the tube bond than for the sheet bond.

**Figure 4. Effect of variable $\Delta G_{Sh} - \Delta G_{Tu}$ (with fixed values of $\Delta G_{Tu} = -15.5\ k_B T$ and $\Delta G_{ShSh} = 6\ k_B T$) on the assembly process.** (a)-(e) plot the simulation results with constant free dimer concentration and (f) plots the results with constant total dimers. (a) Percentage of ribbon structures v.s. time for different values of $\Delta G_{Sh} - \Delta G_{Tu}$ (−2, −1, 0, 1, 2 $k_B T$ as labeled in the figure with corresponding circled numbers). (b) Probability of finding neighboring *tube-sheet* (T-S) structures as a function of time ($\Delta G_{Sh} - \Delta G_{Tu} = -2$, −1, 0, 1, 2 $k_B T$ as labeled in the figure with circled number). (c) Percentage of T-S structures v.s. time for structures with 3 PFs (solid line) and structures with 4 or more PFs (dashed line). (d) Average PF lengths of assembly structures v.s. time with number of PF =1, 2, 3, 4, 5, and 6, respectively (for $\Delta G_{Sh} - \Delta G_{Tu} = 1\ k_B T$). (e) Occurrence of different size clusters v.s. time with numbers of PF = 1, 2, 3, 4, 5 and 6, respectively ($\Delta G_{Sh} - \Delta G_{Tu} = 1\ k_B T$ for all). (f) Percentage of T-S structures v.s. time for variable $\Delta G_{Sh} - \Delta G_{Tu}$ (−2, −1, 0, 1, 2 $k_B T$, as labeled in the figure with corresponding circled numbers) with a constant number of total tubulin dimers of 100.

**Figure 5. Effect of variable $\Delta G_{ShSh}$ on the assembly structures for fixed $\Delta G_{Sh} = -17\ k_B T$ and $\Delta G_{Tu} = -15.5\ k_B T$ ($\Delta G_{Sh} - \Delta G_{Tu} = -1.5\ k_B T < 0$).** (a) Percentage of ribbon structures as a function of time ($\Delta G_{ShSh}$ = 0, 2, 4 and 6 as indicated by circled numbers). (b) Trimer-structure distribution v.s. simulation step for $\Delta G_{ShSh} = 0$. The three possible trimer structures, T-T (*tube-tube*), T-S (*tube*-sheet) and S-S (sheet-sheet), are indicated in the figure. (c) Trimer structure

distribution v.s. simulation step with $\Delta G_{ShSh} = 2\ k_B T$. (d) Trimer-structure distribution v.s. simulation step with $\Delta G_{ShSh} = 6\ k_B T$.

**Figure 6. Effects of tubulin dimer concentrations on the assembly process** (for $\Delta G_{SS} = 6\ k_B T$, $\Delta G_{Sh} = -14\ k_B T$, $\Delta G_{Tu} = -15.5\ k_B T$, i.e., $\Delta G_{Sh} - \Delta G_{Tu} = 1.5\ k_B T$) (dimer concentration c = 125, 25, 5 $\mu$M, as labeled in the figure with corresponding circled numbers). (a) Percentage of ribbon structures as a function of time. (b) Probability of finding neighboring T-S structures as a function of time. (c) Average PF lengths of structures with 1 PF (dashed lines with grey circled numbers indicating concentrations) and 5 PFs (solid lines with open circled numbers indicating concentrations) v.s. time. (d) Occurrence of clusters with 5 PFs v.s. time for different tubulin concentration as labeled.

**Figure 7. Schematic illustration of how two PFs could form sheet bonds fast and then be kinetically trapped.** (a) Illustrative pathways of the assembly process showing a kinetic trap. (b) Schematic illustration that formation of additional sheet bonds increases the transition barrier to the thermodynamically more stable *tube* bonds.

**Figure 8. Schematic illustration of force generation models.** (a) the ratchet model based on the dimer direct-addition model and (b) the possible force generation mechanisms for the new model.

**Table 1. Parameters used in the simulation.**

| Parameters | Values | References |
|---|---|---|
| Longitudinal bond strength extracting part of the entropy term $\Delta G_{long}$ | $-19\ k_B T$ | [4,26][*] |
| Sheet bond strength extracting part of the entropy term $\Delta G_{Sh}$ | Scheme 1: $-13.5 \sim -17.5\ k_B T$, Scheme 2: $-13\ k_B T$ | varying parameter |
| *Tube* bond strength $\Delta G_{Tu}$ | Scheme 1: $-15.5\ k_B T$, Scheme 2: $-16.5\ k_B T$ | [4,26][*] |
| Energy barrier $\Delta G_{lactST}$ | $-9.5\ k_B T$ | estimated |
| Entropy loss for two dimer assemble $\Delta G_{Entropy}(1 \rightarrow 2)$ | $5.5\ k_B T$ [†] | [2,4] |
| Mutual interaction energy for sheet-sheet bonds $\Delta G_{ShSh}$ | Scheme 1: $0 \sim 6\ k_B T$, Scheme 2: $0\ k_B T$ | varying parameter |
| Mutual interaction energy for *tube-tube* bonds $\Delta G_{TuTu}$ | Scheme 1: $0\ k_B T$, Scheme 2: $0 \sim 6\ k_B T$ | varying parameter |
| Rate constant for longitudinal assemble at plus end $k_{long}$ | $2 \times 10^6\ \mu M s^{-1}$ | [4,26,47,48,49,50] |
| Rate constant for longitudinal assemble at minus end $k_{nLong}$ | $k_{Long} \times \delta$ | [12,13] |
| Assemble ratio between minus and plus ends $\delta$ | $1/3$ | [12,13] |
| Rate constant for lateral assemble with *tube* bond $k_{Tu}$ | $5 \times 10^3\ \mu M \cdot s^{-1}$ | [4] |
| Rate constant for lateral assemble with sheet bond $k_{Sh}$ | $1 \times 10^5\ \mu M \cdot s^{-1}$ | estimated |
| Rate constant for conversion between sheet and *tube* bonds $k_{ST0}$ | $5 \times 10^4$ $\mu$Ms$^{-1}$ | estimated |
| Tubulin concentration $c$ | 25 $\mu$M unless specified otherwise | [9] |

* Derived quantities, See Supporting text B for explanation.

†The entropy term for processes other than 1→2 is discussed in Supporting text B.

Figure 1

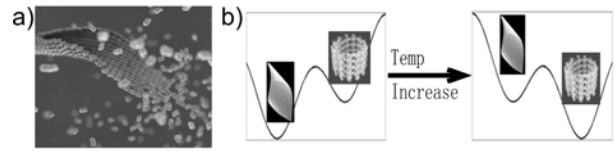

Figure 2

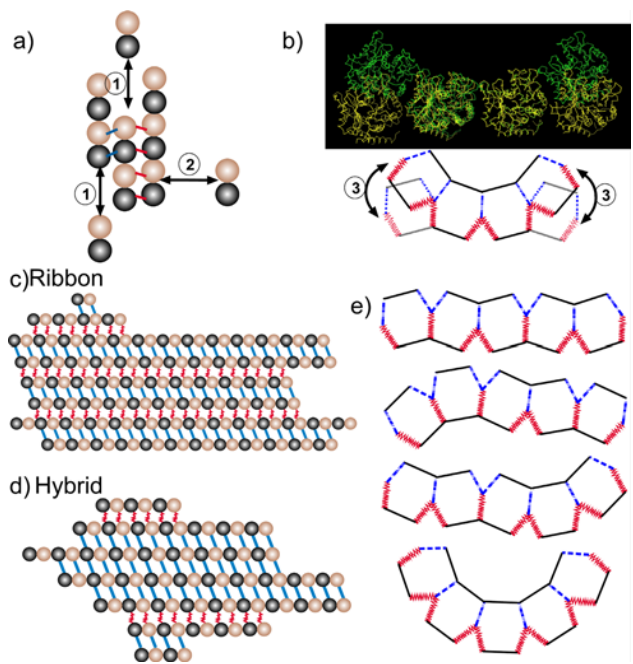

Figure 3

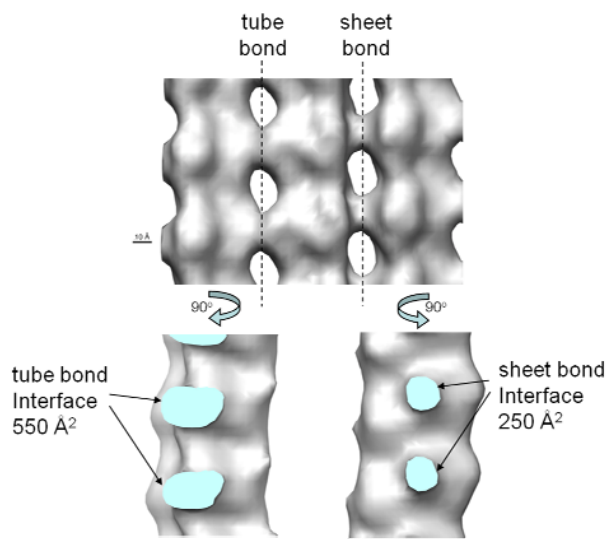

Figure 4

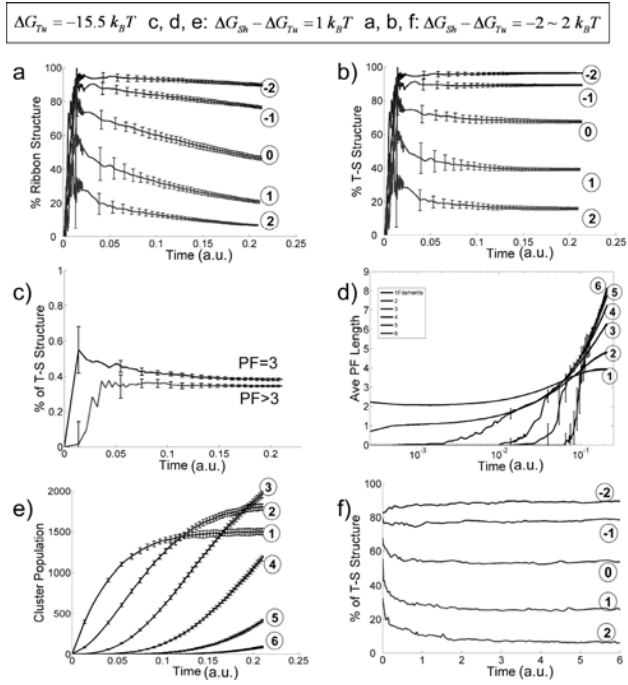

Figure 5

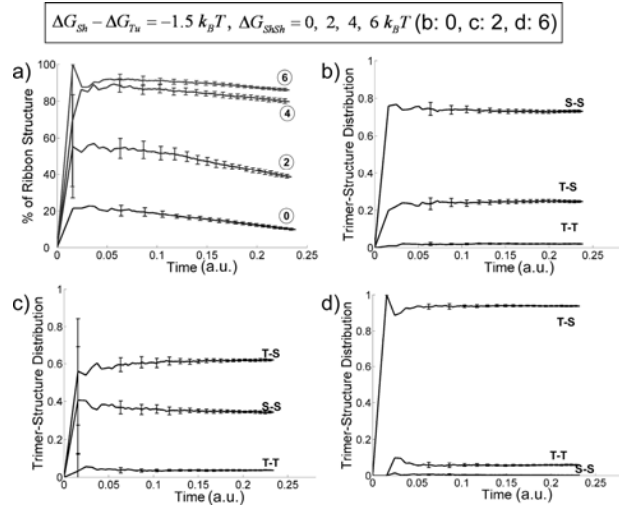

Figure 6

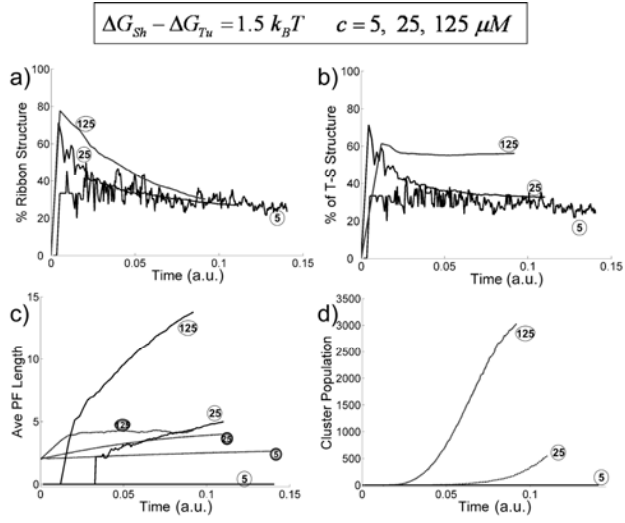

Figure 7

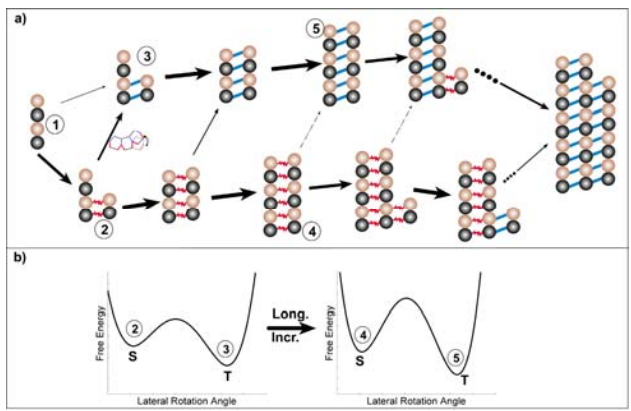

Figure 8

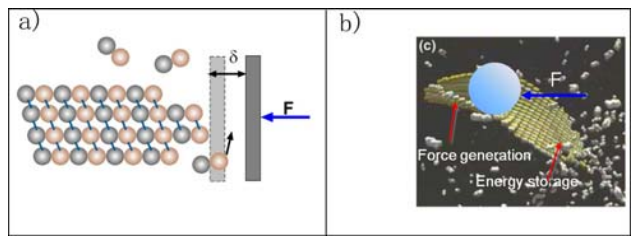

## Supporting text

**A. Rate constants in the model:** As discussed in the main text, three types of basic reactions are considered in our model (Fig. 2a). The following formula give the corresponding reaction rates satisfying detailed balance:

1) Longitudinal On/Off: The reaction rate constants for dimer addtion/dissociation from the plus end of a protofilament are given by

$$k_L^{N \to N+1} = k_L^{1 \to 2}$$

$$k_{-L}^{N+1 \to N} = k_L^{N \to N+1} \exp\left(\left(\Delta G_{long} + i\Delta G_{lat} + \Delta G_{Entropy}(N \to N+1)\right)/k_B T\right)$$

where $k_{-L}$ is the disassociating rate, related to the associating rate $k_L$ by the effective binding free energy. We assume that the association rate is independent of the protofilament length. $\Delta G_{long}$ is the longitudinal bond binding energy, $\Delta G_{lat}$ is the binding energy for a lateral bond, which can be either *tube* bond or sheet bond. The corresponding values are $\Delta G_{Tu}$ and $\Delta G_{Su}$. The number $i$ refers to the number of lateral bonds formed during the association process, which ranges from 0 to 2. $\Delta G_{Entropy}$ accounts for lose of translational and rotational entropy during the association process, which is defined as positive. $\Delta G_{Entropy}(N \to N+1)$ represents the entropic portion of the energy corresponding to adding one dimer onto the cluster with N dimers. For small N, $\Delta G_{Entropy}(N \to N+1)$ also depends on the cluster shape. Its calculation is discussed below. The term $k_B$ is the Boltzmann constant, and $T$ is the temperature in Kelvin. The minus end longitudinal reactions are the same as plus end reactions, except for a constant factor $\delta \sim 0.3$ [1]

2) Lateral On/Off: The rate constants for one dimer lateral association or disassociation from the cluster are

$$k_{-Tu} = k_T \exp\left(\Delta G_{Tu}/k_B T + \Delta G_{Entropy}(1 \to 2)/k_B T\right),$$
$$k_{-Sh} = k_S \exp\left(\Delta G_{Sh}/k_B T + \Delta G_{Entropy}(1 \to 2)/k_B T\right),$$

Where subscript *Tu* stands for *tube* binding, and *Sh* stands for sheet binding. Lateral association rates $k_{Tu}$ and $k_{Sh}$ are determined by comparing existing model parameters and proposed mechanism. $k_{-Tu}$ and $k_{-Sh}$ are disassociation rates. Free energies are defined in the same way as for the longitudinal reaction.

3) Switch of lateral bonds: The conversion rate constants between *sheet* bond and *tube* bond are given by

$$k_{Tu \to Sh}^n = k^0 \exp\left(n\left(\Delta G_{Tu} + \varepsilon \Delta G_{TuTu} - \Delta G^+\right)/k_B T\right),.$$

$$k_{Sh \to Tu}^n = k^0 \exp\left(n\left(\Delta G_{Sh} + \zeta \Delta G_{ShSh} - \Delta G^+\right)/k_B T\right)$$

where the activation energy, $\Delta G^+ > \Delta G_{Tu}, \Delta G_{Sh}$, is the energy barrier between two bond types, $k^0$ is a constant, $\Delta G_{ShSh}$ and $\Delta G_{TuTu}$ are the allosteric terms in Scheme 1 and Shceme 2 (see supporting text C and also in main text), respectively. The parameters $\varepsilon$ and $\zeta$ can assume values 0, 1, or 2, depending on the lateral bond types of neighbor filament pairs. For instance, $\zeta = 0$ if both neighbor filament pairs have tube bonds. The switching rates decrease quickly with increasing number of lateral bonds.

**B. Calculation of the entropic contribution:** Erickson discussed the necessity of treating different free energy contributions, especially the translational and rotational entropy, separately [2]. He discussed the situation adding one tubulin dimer to a large growing microtubule. In our case the system starts with dimers, and form larger and larger clusters. Therefore, we will need to generalize the procedure of Erickson, as discussed below.

The entropic term appears in both longitudinal and lateral reactions. In our model, we consider only the rotational and translational entropic energy. To estimate $\Delta G_{Entropy}$, we consider the partition function of rotational and translational motion of a cluster with N dimers,

$$Q_N = q_N^t q_N^r;$$

where N is the total number of dimers in the cluster. The subscript $t$ stands for translation and $r$ for rotation. The entropy can then be written as

$$S_1 = k_B \log Q_1 + k_B T \frac{\partial}{\partial T} Q_1$$

$$S_N = k_B \log Q_N + k_B T \frac{\partial}{\partial T} Q_N$$

where $S_1$ and $Q_1$ are the entropy and the partition function for one dimer, respectively. We approximate a dimer as a rectangular cuboid with dimensions $h_{height} \times d \times w$. A cluster has a structure (approximately) of a cuboid of dimensions $H \times D \times W$, with $H = \langle n_h \rangle \cdot h_{height}$, $D = d$, and $W = n_w \cdot w$. $\langle n_h \rangle$ is the average number of dimers along the longitudinal direction. $n_w$ is the number of filaments.

Therefore, the partition function can be written as:

$$q_N^t = \left(\frac{2\pi n_w n_h m k_B T}{h^2}\right)^{3/2} V$$

$$q_N^r = \pi^{1/2} \left(\frac{8\pi^2 k_B T}{h^2}\right)^{3/2} \left(I_{hN} I_{wN} I_{dN}\right)^{1/2},$$

where $V$ is the volume, and $h$ is the Planck's constant. The principal moments of inertia for a cuboid structure are

$$I_{hN} = \frac{1}{12} m n_w n_h \left( n_w^2 w^2 + d^2 \right);$$

$$I_{wN} = \frac{1}{12} m n_w n_h \left( n_h^2 h_{height}^2 + d^2 \right);$$

$$I_{dN} = \frac{1}{12} m n_w n_h \left( n_w^2 w^2 + n_h^2 h_{height}^2 \right).$$

Combining all the equations above, we have

$$\Delta G_{Entropy}(N \to N+1) = -T\Delta S_{N,1 \to N+1} = -T(S_{N+1} - S_N - S_1) = F(n_w, \langle n_h \rangle)\Delta G_{Entropy}(1 \to 2)$$

where $F(n_w, n_h)$ is in general a function of $N=n_w n_h$ (derived from the partition functions given above) that represents the entropic energy ratio of adding one dimer to the cluster with $N$ dimers versus adding one dimer to another dimer. Erickson pointed out that calculating $\Delta G_{Entropy}(N \to N+1)$ directly from the corresponding partition function result in overestimation [2,3]. Instead the above relation allows us to link $\Delta G_{Entropy}(N \to N+1)$ to $\Delta G_{Entropy}(1 \to 2)$. The value of $\Delta G_{Entropy}(1 \to 2)$ is obtained by requiring that when $N$ is large, $\Delta G_{Entropy}(N \to N+1)$ tends to a constant value of $\sim 10\ k_B T$, as suggested by Erickson and by Howard [2,3]. The value of the overall binding free energies ($\sim -9\ k_B T$ for longitudinal, and $\sim -5.5\ k_B T$ for lateral tube binding interactions) are close to what used in other model studies [4].

We want to point out that the detailed treatment of the binding energy, especially the entropic term, is not essential for the conclusion made in the main text. However, it makes the model more consistent, since the dependence of entropic change on the cluster size can affect the rates by orders of magnitude [3].

**C. Physical origins of the temperature dependence of the free energy terms:**

1) For Scheme 1, we focus on the temperature dependence of $(\Delta G_{Sh} - \Delta G_{Tu})$. Physically the potential near a stable protein conformation can be approximated as a set of harmonic potentials,

$$V = V_0 + \frac{1}{2} \sum_i \kappa_i x_i^2$$

where $\{\kappa\}$ are spring constants. The harmonic approximation makes the following analysis easy, but is unnecessary for reaching our final conclusion. The corresponding classical partition function (we neglect quantum effects which don't change the result qualitatively here) is

$$Q = \int e^{-\frac{1}{k_B T}\left(V_0 + \sum_i \frac{1}{2}\kappa_i x_i^2\right)} dx_1 dx_2 ... dx_N = e^{-\frac{1}{k_B T}V_0} \prod_i \left( \frac{2\pi k_B T}{\kappa_i} \right)^{1/2}$$

The free energy is

$$G = -k_B T \ln Q$$

$$= -\frac{1}{2} k_B NT \ln T + k_B T \ln \left( \prod_i \left( \frac{\kappa_i}{2\pi k_B} \right)^{1/2} \right) + V_0$$

$$= -\alpha T \ln T + \beta(\kappa) T + V_0$$

and the free energy difference $\Delta G_{Sh} - \Delta G_{Tu} = (V_{S0} - V_{M0}) + (\beta_{Sh} - \beta_{Tu})T$, where $\alpha$ and $\beta(\kappa)$ are positive. Therefore it is possible that $(\Delta G_{Sh} - \Delta G_{Tu})$ changes sign on increasing temperature, as shown schematically in Fig. S3a.

The sign change of $(\Delta G_{Sh} - \Delta G_{Tu})$ upon increasing temperature implies the entropy change $(\Delta S_{Sh} - \Delta S_{Tu} < 0)$. Another possible source of entropy change is through liberation of water molecules initially bound to protein surfaces. When two protein surfaces interact, some water molecules initially constrained to the surfaces are released to the solution. This can be a huge contribution to entropy increase. Our cryo-EM images reveal more extensive contact surface for the tube bond than for the sheet bond (see Fig. 3). Therefore one might expect more water molecules released upon the tube bond formation than the sheet bond formation, which contributes the relation $(\Delta S_{Sh} - \Delta S_{Tu} < 0)$. Structures at higher resolution will aid in evaluating this hypothesis. With current information, we cannot provide further quantitative analysis.

2) In Scheme 2 we assume that some conformational change (the allosteric effect) accompanies formation of two neighboring lateral bonds. Let's denote the reaction coordinate linking the initial and final conformations $s$. The potential part of the reaction path Hamiltonian [5] along $s$ can be written in the classical form

$$V(s) = V_0(s) + \sum_i \frac{1}{2} \kappa_i(s) x_i^2$$

The classical partition function for the potential of mean force is given by

$$Q(\bar{s}) = \int e^{-\frac{1}{k_B T} V(s)} \delta(s - \bar{s}) \prod_i dx_i ds = \exp\left(-\frac{1}{k_B T} V_0(s)\right) \prod_i \left( \frac{2\pi k_B T}{\kappa_i(\bar{s})} \right)^{1/2}$$

Therefore the free energy change due to the allosteric effect induced conformational change $\Delta G \equiv G(s_b) - G(s_0) = -k_B T (\ln Q_b - \ln Q_0)$ is in the form $\Delta G = V_0(s_b) - V_0(s_0) + (\beta(s_b) - \beta(s_0))T$. The term $\beta$ is defined similarly to what in part 1, except here $\beta$ is dependent on conformational coordinate $s$. If $\beta_{s_b} < \beta_{s_0}$, the allosteric effect $\Delta G_{TuTu}$ decreases as temperature increase (see Fig. S3b). For simplicity we assume that only the allosteric interaction between two consecutive lateral *tube* bonds is appreciable, although the model can be easily generalized. The simulation results based on this scheme are shown in Fig. S4. The figure shows the percentage of ribbon structures at different values of $\Delta G_{TuTu}$. As discussed above, the different values of $\Delta G_{TuTu}$ correspond to different temperatures. The figure shows that smaller $\Delta G_{TuTu}$ values (higher temperature) give lower percentage of ribbon structures than larger $\Delta G_{TuTu}$ values do. For $\Delta G_{TuTu} = 6k_BT$, the clusters contain over 90% ribbon structures, compared to the 10% for $\Delta G_{TuTu} = 0k_BT$. The results indicate that Scheme 2 is a good alternative explanation to the existing experimental data. To discriminate between Schemes 1 and 2, more data, especially the structures with 2 PFs, would be needed.

We want to point out that entropy is the primary driving force for many biological processes, e.g., hydrophobic interactions. It is physically reasonable that the entropy term leads ($\Delta G_{Sh} - \Delta G_{Tu}$) to change its sign upon temperature change, especially if $\Delta G_{Sh}$ and $\Delta G_{Tu}$ are very close, as what we used in this work. Experimentally we found that at physiological magnesium concentration, GMPCPP tubulins form normal microtubule structure at 37°C, but only short single PF structures at lower temperature. These observations are consistent with our assumption that entropy has large contribution to the lateral bond energies. Increasing the temperature stablizes both types of the lateral bonds, which, esp. the sheet bond, can be further stabilized by increasing the magnesium concentration. Alternatively, one may suggest a kinetic explanation for the lacking of larger structures at lower temperature: the lateral bond formation rates are too slow. However, no larger structure is observed at longer time (in hours). This observation doesn't support the kinetic explanation.

1. Summers K, Kirschner MW (1979) Characteristics of the polar assembly and disassembly of microtubules observed in vitro by darkfield light microscopy. J Cell Biol 83: 205-217.
2. Erickson HP (1989) Co-operativity in protein-protein association : The structure and stability of the actin filament. J Mol Biol 206: 465-474.
3. Howard J (2001) Mechanics of Motor Proteins and the Cytoskeleton. Sunderland, MA: Sinauer.
4. VanBuren V, Cassimeris L, Odde DJ (2005) Mechanochemical Model of Microtubule Structure and Self-Assembly Kinetics. Biophys J 89: 2911-2926.
5. Miller WH, Handy NC, Adams JE (1980) Reaction path Hamiltonian for polyatomic molecules. J Chem Phys 72: 99-112.
6. Wang H-W, Nogales E (2005) Nucleotide-dependent bending flexibility of tubulin regulates microtubule assembly. Nature 435: 911-915.
7. Fygensonm D, Needleman D, Sneppen K (2004) Variability-based sequence alignment identifies residues responsible for functional differences in α and β tubulin. Protein Sci 13: 25-31.
8. Sui H, Downing KH (2006) Molecular architecture of axonemal microtubule doublets revealed by cryo-electron tomography. Nature 442: 475-478.

## Supporting Figure Legends

**Figure S1. Structural basis for the two types of lateral bonds.** (a) Structure of the $\alpha\beta$-tubulin dimer with residues involved in lateral interactions indicated. Blue: residues engaged in lateral *tube* bonds (274-286, 52-61). Red: residues engaged in lateral *sheet* bond (336-342, 158-164) (these residues have been identified by docking the high-resolution tubulin structure into the 18 Å reconstruction of the ribbon [6], and therefore are correct within the constrains of the limited resolution). Pink and yellow: possible surface residues (108-130, 209-225, 300-311) along the *tube*-sheet conversion pathway. (b) Variability-based sequence alignment of $\alpha-$ and $\beta-$ tubulin performed by Fygenson et al. [7]. The blue and red boxes indicate the residues involved in the *tube* and *sheet* bond formation given in (a), respectively. The figure is adapted from Fig. 2 of Fygenson et al. [7] with permission. (c) Comparison of the non-MT lateral interactions observed in the microtubule doublet of axonemes (top) [8] (PDB file provided by Sui and Downing) and the ribbon structures (bottom) [6].

**Figure S2. Effect of variable $\Delta G_{ShSh}$ on the assembled structures with $\Delta G_{Sh} = -14.5\ k_BT$ and $\Delta G_{Tu} = -15.5\ k_BT$ ($\Delta G_{Sh} - \Delta G_{Tu} = 1\ k_BT > 0$).** The figure shows the percentage of ribbon structures as a function of the time for $\Delta G_{ShSh} = 0, 1, 2$ and $3\ k_BT$, as indicated.

**Figure S3. Schematic Illustration of the physical origins of the temperature dependence of the free energy terms.** (a) $\Delta G_{Sh}$ and $\Delta G_{Tu}$ have different temperature dependence and their difference changes sign over $T$. (b) The dependence of $\Delta G_{TuTu}$ on the conformational coordinate describing the necessary collective conformational change upon forming two neighboring lateral *tube* bonds varies with temperature.

**Figure S4. Effects of variable $\Delta G_{TuTu}$ on the assembly structures using the Scheme 2 described in Fig. S3b.** (0, 2, 4, and $6\ k_BT$, as indicated by corresponding circled numbers). Different $\Delta G_{TuTu}$ correspond to different temperatures as showed in Fig. S3b and supporting text C. $\Delta G_{Sh} = -13\ k_BT$ and $\Delta G_{Tu} = -16.5\ k_BT$ were used for all simulations. Other parameters are the same as in the Scheme 1 described in detail in the main text. The final results are averaged over 60 independent simulations. (a) Percentage of ribbon structure v.s. simulation step. (b) Percentage of T-S structure. (c) Average PF length for clusters of different size (1 to 6 PFs as indicated by circled numbers), with $\Delta G_{TuTu} = 2\,k_B T$. (d) Cluster population for clusters of different size (1 to 6 PFs as indicated by circled numbers), with $\Delta G_{TuTu} = 2\,k_B T$.

**Figure S5. Population ratio of *tube*-cluster versus *sheet*-cluster for 2-PF structures as a function of time.** Solid and dashed lines with triangles correspond, respectively, to Scheme 1 ($\Delta G_{ShSh} > 0$, $\Delta G_{TuTu} \sim 0$, $\Delta G_{Sh} - \Delta G_{Tu} = 1.5\,k_B T$, $\Delta G_{ShSh} = 6\,k_B T$) and to Scheme 2 ($\Delta G_{TuTu} > 0$, $\Delta G_{ShSh} \sim 0$, $\Delta G_{Sh} - \Delta G_{Tu} = 3.5\,k_B T$, $\Delta G_{TuTu} = 2\,k_B T$), both at high temperature. The lines without triangles are for Scheme 1 (solid line, $\Delta G_{Sh} - \Delta G_{Tu} = -1.5\,k_B T$. $\Delta G_{ShSh} = 6\,k_B T$) and Scheme 2 (dashed line, $\Delta G_{Sh} - \Delta G_{Tu} = 3.5\,k_B T$ $\Delta G_{TuTu} = 6\,k_B T$) at low temperature.

Figure S1

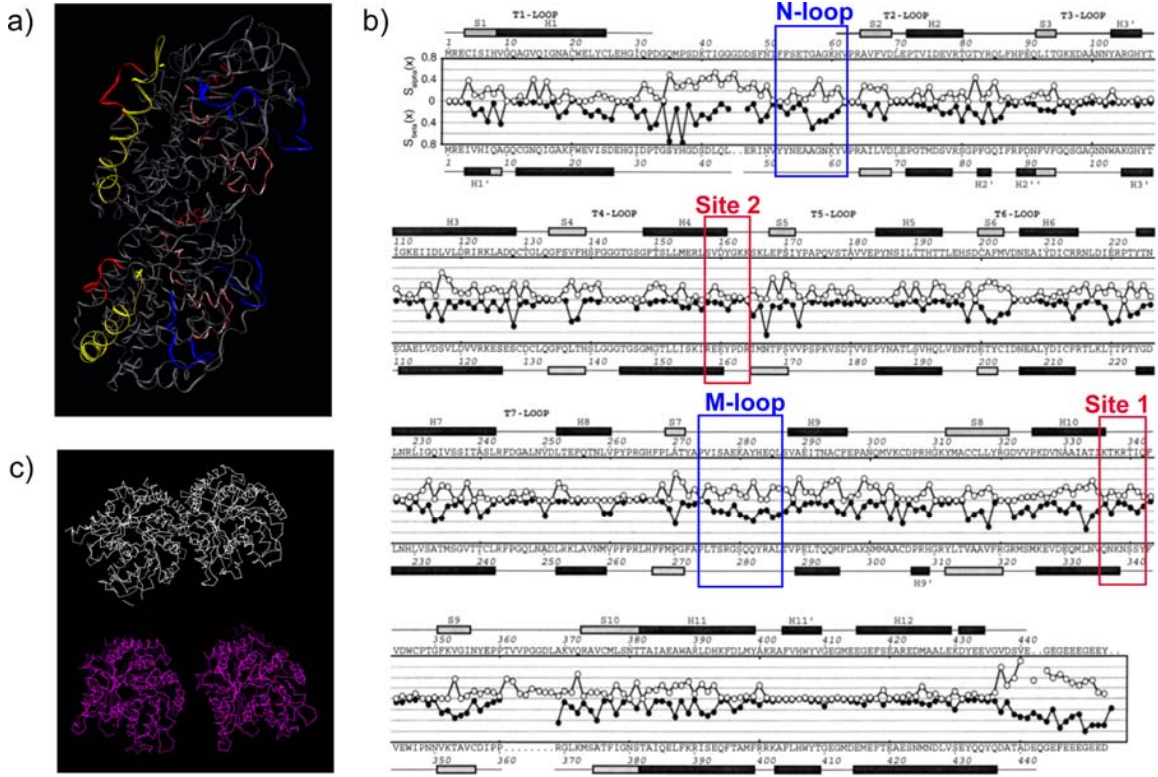

Figure S2

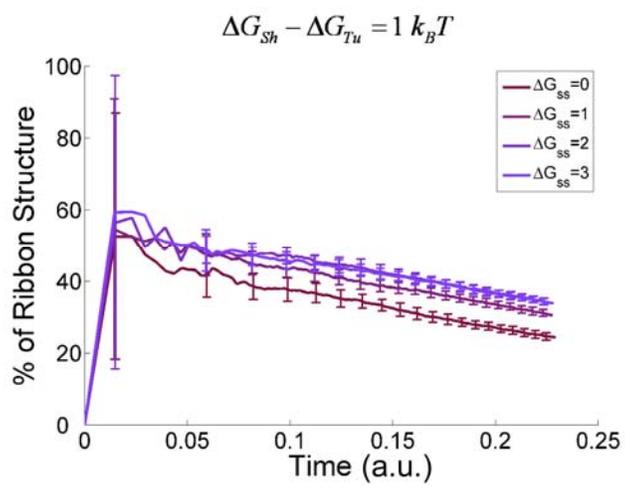

Figure S3

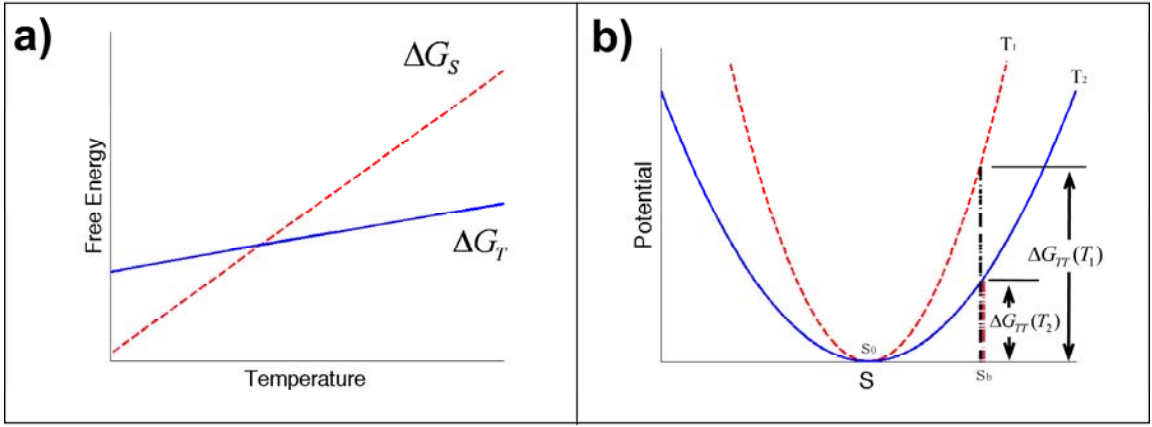

Figure S4

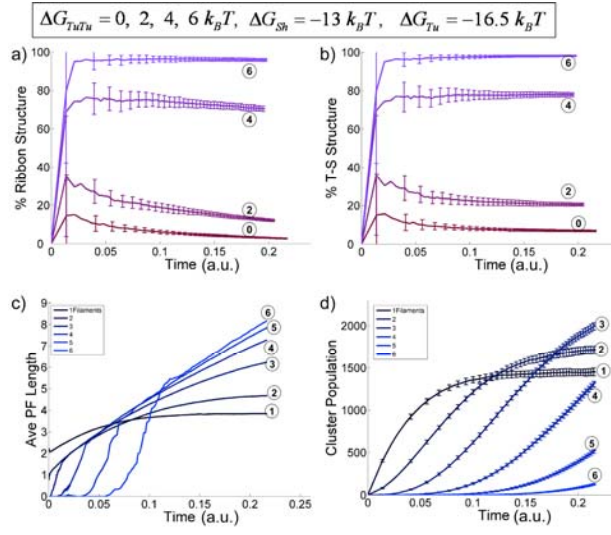

Figure S5

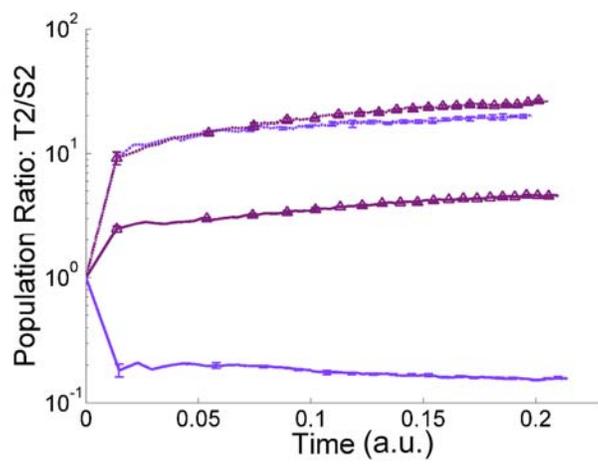